\documentclass[aps,pra,showpacs,twocolumn,amsmath,amssymb,superscriptaddress]{revtex4}
\usepackage[english]{babel}
\usepackage{latexsym}
\usepackage{graphicx}
\usepackage{epsf} 
\usepackage{color}
\usepackage{url}
\usepackage[normalem]{ulem}
\usepackage{amsthm}
\usepackage{amsmath}
\usepackage[utf8]{inputenc}
\usepackage[T1]{fontenc}
\usepackage{epsfig}

\usepackage{graphicx}

\def\be{\begin{equation}}
\def\ee{\end{equation}}
\def\bea{\begin{eqnarray}}
\def\eea{\end{eqnarray}}
\def\bi{\begin{itemize}}
\def\ei{\end{itemize}}
\def\bin{\begin{enumerate}}
\def\ein{\end{enumerate}}

\begin{document}
\title{Localization in random fractal lattices}


\author{Arkadiusz Kosior}
\affiliation{Instytut Fizyki imienia Mariana Smoluchowskiego, Uniwersytet Jagiello\'nski, \L{}ojasiewicza 11, 30-348 Krak\'ow, Poland}
\affiliation{ICFO-Institut de Ciencies Fotoniques, Av. Carl Friedrich Gauss, 3, 08860 Castelldefels (Barcelona), Spain}
\author{Krzysztof Sacha}
\affiliation{Instytut Fizyki imienia Mariana Smoluchowskiego, Uniwersytet Jagiello\'nski, \L{}ojasiewicza 11, 30-348 Krak\'ow, Poland}
\affiliation{Mark Kac Complex Systems Research Center, Uniwersytet Jagiello\'nski, \L{}ojasiewicza 11, 30-348 Krak\'ow, Poland}

\date{\today}

\begin{abstract}
We investigate the issue of eigenfunction localization in random fractal lattices embedded in two dimensional Euclidean space. In the system of our interest, there is no diagonal disorder -- the disorder arises from random connectivity of non-uniformly distributed lattice sites only. By adding or removing links between lattice sites, we change the spectral dimension of a lattice but keep the fractional Hausdorff dimension fixed. From the analysis of energy level statistics obtained via direct diagonalization of finite systems, we observe that eigenfunction localization strongly depends on the spectral dimension. Conversely, we show that localization properties of the system do not change significantly while we alter the Hausdorff dimension. In addition, for low spectral dimensions, we observe superlocalization resonances and a formation of an energy gap around the center of the spectrum.

\end{abstract}


\maketitle
 \section{Introduction}

Anderson localization (AL) is a single-particle disorder induced effect which leads to exponential localization of particles' eigenfunctions \cite{Anderson1958,Krame1993,VanTiggelen1999,MuellerDelande:Houches:2009,Lagendijk2009}. In his groundbreaking work, Anderson considered non-interacting electronic gas in a tight-binding model in the presence of on-site disorder. Since then, AL was investigated in many different models, including off-diagonal disorder \cite{Eilmes1998,Reza2010,Biddle2011}, disorder correlations \cite{Moura1998,Piraud2013,Kosior2015,Major2016}, random fluxes \cite{Kalmeyer1993,Sheng2000}, localization in the momentum space of classically chaotic systems \cite{Fishman,Garreau16} and, recently, localization in the time domain \cite{sacha2015,sacha2016}. 

The interest in AL renewed after the first experimental observation of the phenomenon in ultracold atomic gases \cite{Billy2008,Aspect2009,Modugno2010,Inguscio2008}. Although the Anderson model was created to describe electronic gases, AL is difficult to observe in metals due to electron-phonon and electron-electron interactions. On the contrary, interatomic interactions can be switched off  in a system of ultracold atomic gases trapped in optical lattice potentials \cite{Chin2010}. An optical lattice serves as an artificial, phononless crystalline structure,  whose geometry and properties can be easily changed \cite{Jaksch2005,Bloch2008}. Therefore, in recent years,  ultracold atoms in optical lattices have become a very important toolbox used to test diverse physical models and phenomena \cite{feynmann1982,Buluta2009,Hauke2012}. 

The dimensionality of a system plays an important role in the context of AL \cite{Krame1993,VanTiggelen1999,Lagendijk2009}. In particular, in three dimensional (3D) space a  phase transition occurs at a critical energy, called the mobility edge, separating localized and extended states \cite{Abrahams,Mott1987}. It is therefore natural to investigate AL phenomenon  in systems with non-integer dimension, i.e. in fractals \cite{Nakayama1994,Garcia2010}. Whereas in Euclidean space it is sufficient to define one kind of space dimension, in the case of fractals one needs to distinguish: the dimension of the embedding Euclidean space~D, the Hausdorff dimension $d_H$ and the spectral dimension $d_s$ \cite{Nakayama1994,Rammal1983}. Whereas the Hausdorff dimension describes how the number of sites scales with the system size, the spectral dimension is related to a random walk on the lattice: the number of distinct sites $S_n$ visited by a random walker in $n$ steps scales as $S_n \propto n^{d_s/2}$, provided that $d_s < 2$. The studies proved that it is the spectral dimension $d_s$ that is relevant in AL and that $d_s=2$ is the lower critical dimension, below which all the states are localized in the presence of a disorder potential \cite{Schreiber1996}.   

\begin{figure}[bt]
\begin{center}
\resizebox{0.9\columnwidth}{!}{\includegraphics{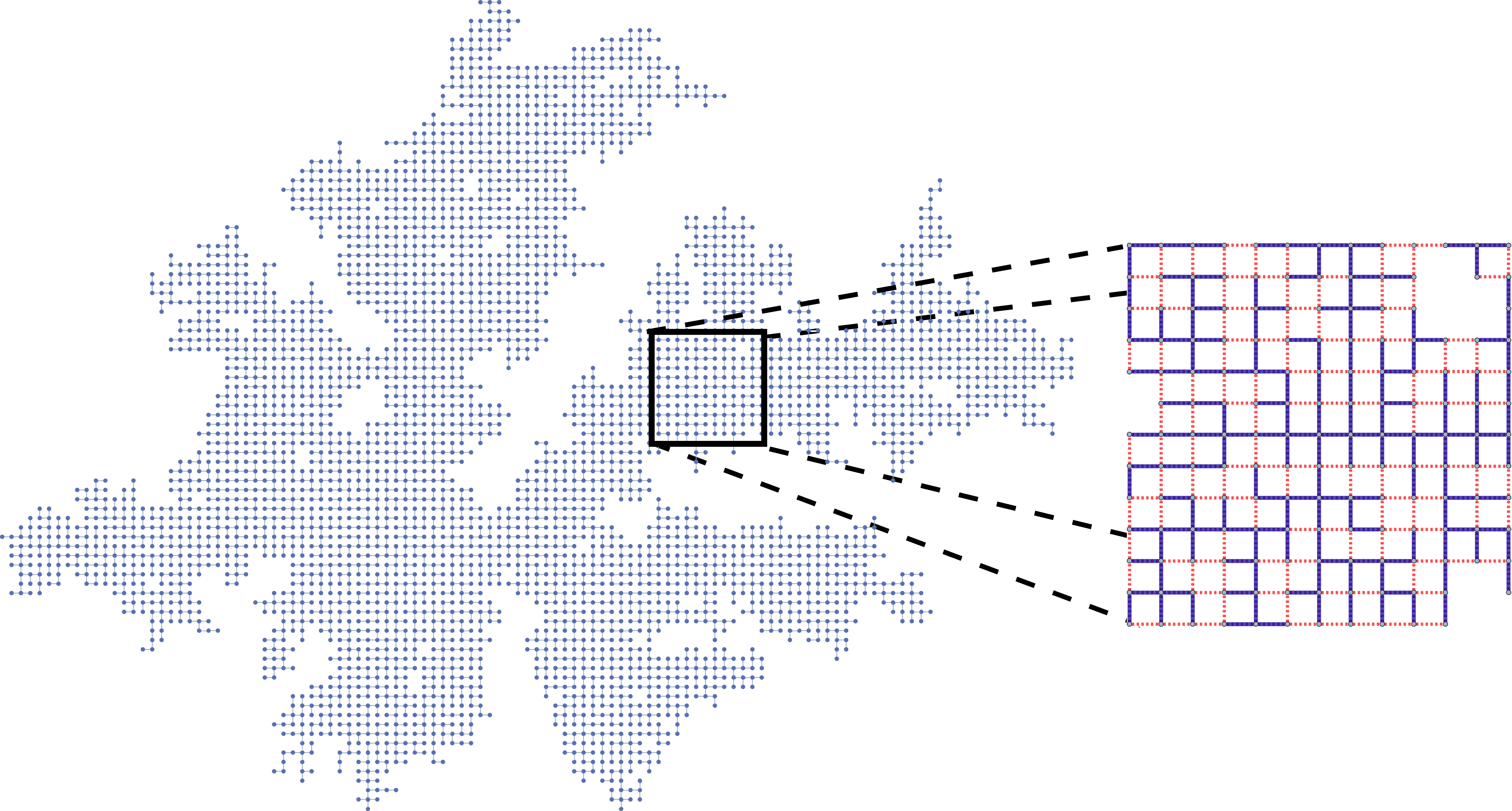}}
\caption{(color online) An example of a minimal random fractal lattice (RFL) mapped on the 2D square lattice for $\eta=1$, see text.
The solid blue lines indicate that the neighboring sites are linked, i.e. the quantum tunneling between these sites is possible. Red dotted lines represent the lack of quantum tunneling between nearest neighbors. By adding links to the lattice (i.e. replacing a number of red dotted lines with blue lines) one increases the spectral dimension but leaves the Hausdorff dimension unchanged. 
}
\label{fractal_pic}
\end{center}
\end{figure}

 Another important theoretical model in study of localization properties is the quantum percolation model (QP) \cite{Kirkpatrick1972,Nakanishi2009,Schubert2005,Schubert2009}. Anderson model describes particles in the presence of potential disorder (purely diagonal), whilst QP involves the binary kinetic disorder (purely off-diagonal). In QP models the disorder comes from random geometry: a QP lattice, which is a subset of the D-dimension lattice, arises by a removal of a number of sites or links with a probability $q$ (being the only parameter of the model). Despite its simplicity, QP models still arise controversies. The main concern is the question of existence of the localization-delocalization transition in 2D models for $q>0$ (see \cite{Islam2008,Gong2009,Dillon2014} and references therein). In particular, this issue might be important in the context of application of QP models in the description of transport properties in e.g.  manganite films \cite{Zhang2002}, granular metals \cite{Feigelman2004} and doped semiconductors \cite{Drchal2004}.

 In the present paper we investigate eigenfunction localization in random fractal lattices (RFL), i.e. in fractal objects with random site connectivity in the absence of any diagonal disorder, Fig.\ref{fractal_pic}. We would like to stress that QP systems do not belong to fractal objects. That is, in the QP case only an infinite cluster of a percolated lattice at the percolation threshold is a fractal object \cite{Christensen2005}.  Here, on the contrary, we consider a family of lattices with well defined Hausdorff dimension $d_H$ \cite{Niemeyer1984}. Starting with the minimal, connected lattice (i.e. a lattice without loops) and  adding links between nearest neighbors we can increase the spectral dimension $d_s$ of a lattice and keep the Hausdorff dimension $d_H$ fixed. Therefore, the main focus of this paper is to investigate the presence or absence of localization in RFLs while changing the spectral and Hausdorff dimensions independently.  It is worth noting  that the theoretical model investigated here can be realized in ultra-cold atoms laboratories where lattice geometry can be nearly arbitrarily shaped \cite{Bakr2009,Gaunt2012,Bijnen2015}. 

This paper is organized as follows. In Section~\ref{secii} we describe the growth algorithm of random fractal lattices and how the spectral dimension changes  when new lattice links are created. In Section~\ref{seciii} we focus on localization properties of RFLs  and on their dependence on the spectral and Hausdorff dimensions. Particularly, we analyze superlocalization resonances and formation of an energy gap which emerges in the system for small spectral dimension. In Section~\ref{seciv} we  investigate transmission probabilities through the system and quantum evolution of initially localized particles. Finally, in Section~\ref{secv} we conclude.

 \section{Random fractal lattices}\label{secii}

\begin{figure}[tb]
\begin{center}
\resizebox{0.8\columnwidth}{!}{\includegraphics{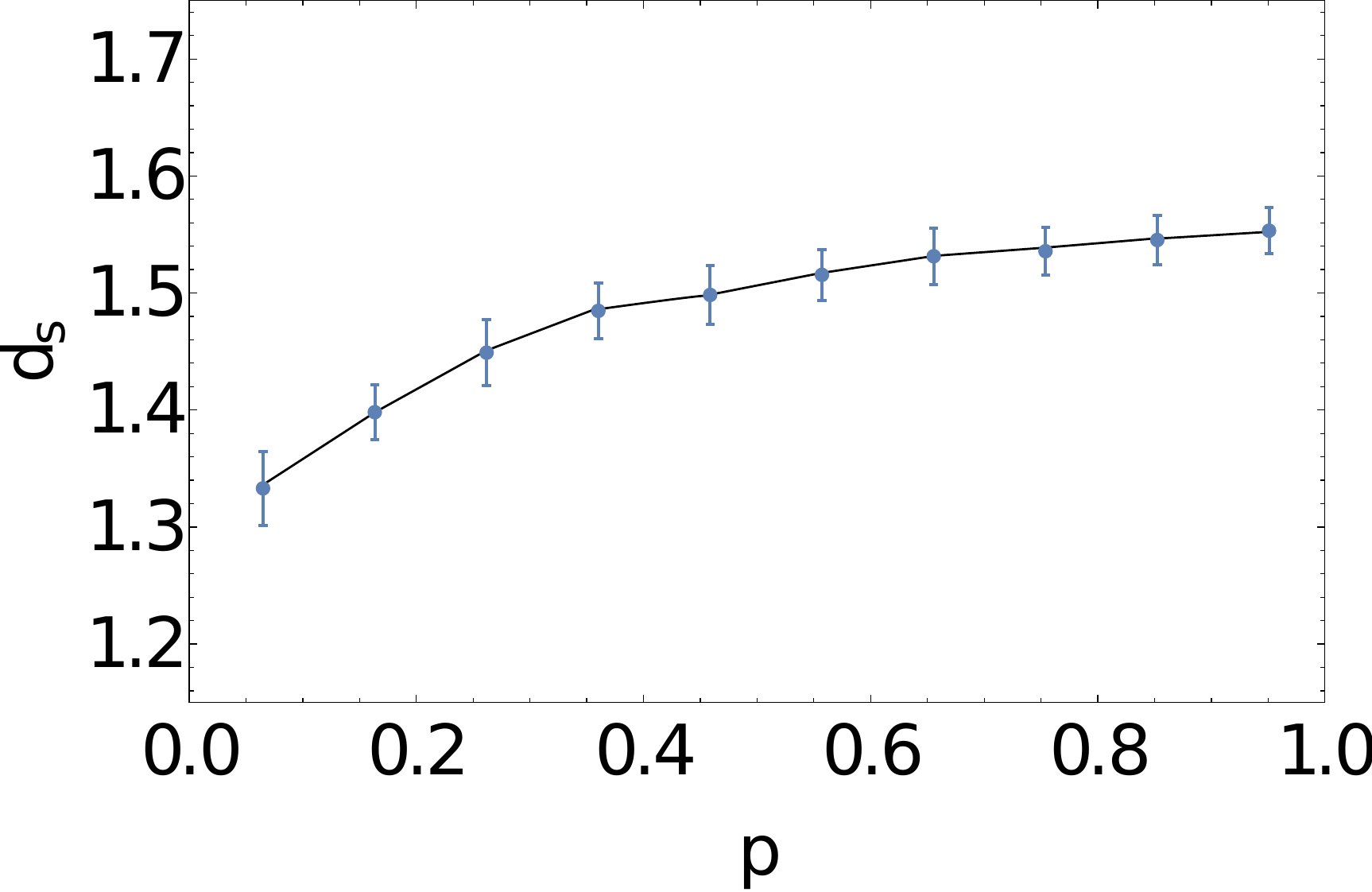}}
\caption{The plot illustrates the change of the spectral dimension of RFLs for the Hausdorff dimension $d_H=1.75\pm0.02$ when increasing the number of lattice links. The dimensionless parameter $p$ is the fraction of links added to the minimal, single connected lattice. A parameter value $p=0$ represents RFLs with the minimal number of links, i.e. no link can be removed without disconnecting a part of the lattice. Conversely, $p=1$ represents RFLs with the maximal number of links, i.e. no link can be added without creating an extra lattice site. The value of the spectral dimension $d_s$ for a given $p$ was obtained from the exponent of the number of distinct sites visited by a random walker $S_n \propto n^{d_s/2}$, and averaging over $2000$ independent RFLs and $500$ realizations of random walks of length $2500$.}
\label{spectral_dim}
\end{center}
\end{figure}

We consider a family of lattices that we call \emph{random fractal lattices} (RFLs), which first arose in a model of dielectric breakdown \cite{Niemeyer1984}. The RFLs are lattices with random site connectivity and with well-defined Hausdorff $d_H$ and spectral $d_s$ dimensions. The Hausdorff dimension, or the capacity dimension, describes how the number of sites scales with the system size. In other words, if a lattice  in Fig.~\ref{fractal_pic} is a fractal object, then the number of sites inside a sphere o radius $r$ is proportional to $r^{d_H}$, where in general $d_H$ is a noninteger exponent \cite{Christensen2005}. On the other hand, the spectral dimension is related to a random walk  on the lattice.  The number of distinct sites $S_n$ covered in $n$ steps of a random walk is proportional to $n^{d_s/2}$, if $d_s < 2$. From a general analysis, the spectral dimension is never larger than the Hausdorff dimension \cite{Nakayama1994,Rammal1983}. 

RFLs under consideration are embedded in the 2D Euclidean space and have their Hausdorff and spectral dimensions smaller than 2. A minimal RFL, i.e. a lattice with the smallest number of lattice links, is a single connected lattice generated by the growth algorithm defined in  Ref.~\cite{Niemeyer1984}. In a nutshell, a new lattice site $(i',j')$ is chosen and linked to the existing lattice at site $(i,j)$  with the probability
\be\label{probability}
P\Big((i,j)\rightarrow (i',j') \Big)= \frac{(\phi_{i',j'})^\eta}{\sum(\phi_{i',j'})^\eta},
\ee
where the summation goes over all possible choices, $\eta$ is a free parameter and $\phi_{i,j}$ is a function fulfilling discrete Laplace equation:
\be
\phi_{i,j}=\frac{1}{4}\left(\phi_{i,j+1}+\phi_{i,j-1}+\phi_{i+1,j}+\phi_{i-1,j}\right).
\label{laplace}
\ee
At start, we set $\phi_{i,j}=0.5$ everywhere. When a site $(i',j')$ is being connected to the lattice, then the value of $\phi_{i',j'}$ is changed to zero. Before linking  another lattice site, the values of $\phi$ in the neighborhood of $(i,j)$ needs to be updated [typically in 5-20 iterations of Eq.~(\ref{laplace})]. The algorithm is stopped after reaching  $N$ lattice sites. The lattices grown in this manner have nonuniform geometry, both in the lattice sites and lattice links occurrence, as in Fig.~\ref{fractal_pic}. In particular, the two neighboring lattice sites do not necessarily need to be connected and the closed loops are forbidden, i.e. a minimal RFL is created. By adding links between nearest neighbors to a given minimal RFL, one opens up new possibilities for a random walker to explore and therefore increases the spectral dimension $d_s$ of the system while keeping the Hausdorff dimension $d_H$ intact. 

The Hausdorff dimension of RFLs depends on the value of the parameter $\eta$ \cite{Niemeyer1984}. For example,  setting $\eta=1$ one can generate a minimal RFL with the Hausdorff dimension $d_H=1.75\pm0.02$ and the spectral dimension $d_s=1.33\pm0.03$. Adding links to a minimal RFL results in an  increase of the spectral dimension, Fig.~\ref{spectral_dim}. At the same time the Hausdorff dimension $d_H$ remains unchanged.

 \section{Localization properties}\label{seciii}

In the following we analyze solutions of the Schr\"odinger equation
\be
E\psi_{(i,j)}=-\sum_{i',j'}\psi_{(i',j')}, 
\label{schrod}
\ee
where $(i,j)$ denotes a position on a RFL embedded in the 2D Euclidean space and the sum runs over nearest neighbor sites if there is a link between $(i,j)$ and $(i',j')$. We assume that the tunneling amplitudes of a particle between neighboring sites and the Planck constant are equal to unity.

\subsection{Analysis of energy level statistics}

In our analysis we investigate  two distinct scenarios:
\begin{itemize}
 \item First, we fix the Hausdorff dimension by setting $\eta=1$ in \eqref{probability}, which corresponds to $d_H=1.75\pm0.02$. Then, we change the spectral dimension in the range between 1.33 and 1.55 by adding links to the minimal RFLs, see Fig.~\ref{spectral_dim}.
\item   In the second scenario, we do the opposite, i.e. we fix the spectral dimension ($d_s\approx1.35$ or $d_s\approx1.5$) and change the Hausdorff dimension by varying the parameter $\eta$ in \eqref{probability}.
\end{itemize}

\begin{figure}[t]
\begin{center}
\resizebox{0.85\columnwidth}{!}{\includegraphics{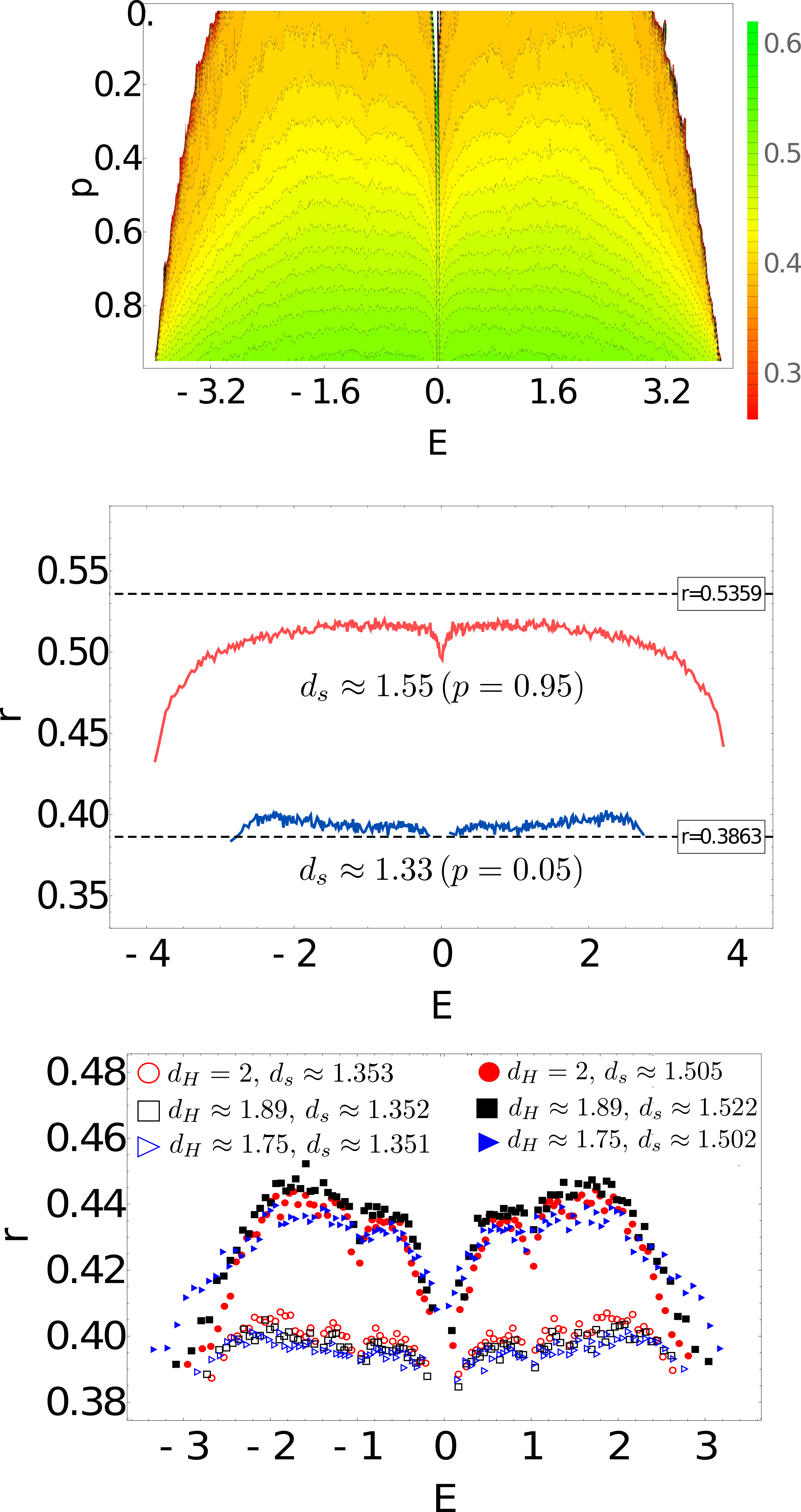}}
\caption{(color online) A plot of the averaged ratio of consecutive energy level spacings $r(E)$ of RFLs with the fixed Hausdorff dimension $d_H=1.75\pm0.02$ (top panel), and two cuts for the extreme cases (middle panel). The vertical axis in the top panel shows the impact of the spectral dimension $d_s$, through the dimensionless parameter $p$, see Fig.~\ref{spectral_dim}, on the level statistics. With increasing $d_s$ the system gradually delocalizes.  For low values of $p$ (corresponding to $d_s\approx 1.35$) and near $E=0$ there is a narrow energy gap emerging, see the discussion in the main text.  The localization is not much influenced by a change of the Hausdorff dimension (bottom panel). The RFL systems with different values of $d_H$ and very similar $d_s$  possess similar localization properties, showing that it is the spectral dimension $d_s$ that is the relevant dimension in this context. All RFLs analyzed here consist of $N=5000$ sites. 
}
\label{phase_diag}
\end{center}
\end{figure}

In order to explore the presence or absence of localization of eigenfunctions of a particle in RFLs, we use a convenient method engaging the energy level statistics obtained via direct diagonalization of finite systems \cite{Oganesyan2007,Pal2010,Tang2015}. Since the localized states usually have very small overlaps (as they may be localized in different parts of the system), the energy levels can be nearly degenerate.  This is why the localization of eigenstates can be observed directly from the spectrum. Therefore, we expect that in the localized phase the energy level statistics follow the Poisson distribution, and in the delocalized phase fulfill the Wigner Dyson distribution \cite{Oganesyan2007,Pal2010,Tang2015}. Having the ordered spectrum of energy levels $\{E_i\}$, we can calculate a quantity
\be\label{ri_eq}
r_i = \frac{\min(\delta_i,\delta_{i-1})}{\max(\delta_i,\delta_{i-1})},
\ee
where $\delta_i=E_i-E_{i-1}$. Next, we average the results over 2000 realizations  of RFL and over neighboring energies $r(E)=\langle r_i \rangle$. The distinction between the two regimes is possible since for the Poisson distribution $r(E)\approx 0.3863$  localized phase) and for the Wigner Dyson distribution $r(E) \approx 0.5359$ (delocalized phase) \cite{Oganesyan2007,Pal2010,Tang2015}. 

To investigate the localization properties  of wavefunctions under the change of the spectral dimension of RFLs, we plot $r(E)$ in Fig.~\ref{phase_diag} (top panel). The vertical axis expresses the spectral dimension via the dimensionless parameter $p$, see Fig.~\ref{spectral_dim}. The panel illustrates strong dependence of the localization properties on the spectral dimension $d_s$. While increasing the spectral dimension we observe a smooth transition from the localized to delocalized phase, what is evident in the middle panel of Fig.~\ref{phase_diag}, where the averaged ratio $r(E)$ is plotted for two extreme values of $d_s$. Furthermore, we  observe nonmonotonous dependence of $r(E)$ on energy (i.e. the localization is stronger near the edges and the center of the spectrum), which has been also observed in QP models \cite{Schubert2005,Gong2009}. 

In the bottom panel of Fig.~\ref{phase_diag} we present  $r(E)$ for a few different Hausdorff dimensions $d_H$ and very similar spectral dimension $d_s$. The data with different $d_H$ do not differ significantly apart from the  edges  of the spectra. The results  show that, similarly to the AL models  \cite{Schreiber1996},  it is the spectral dimension $d_s$ that is the relevant dimension in  the  context of the localization properties of the system. 

\subsection{Energy gap and superlocalization resonances}

\begin{figure}[tb]
\begin{center}
\resizebox{0.8\columnwidth}{!}{\includegraphics{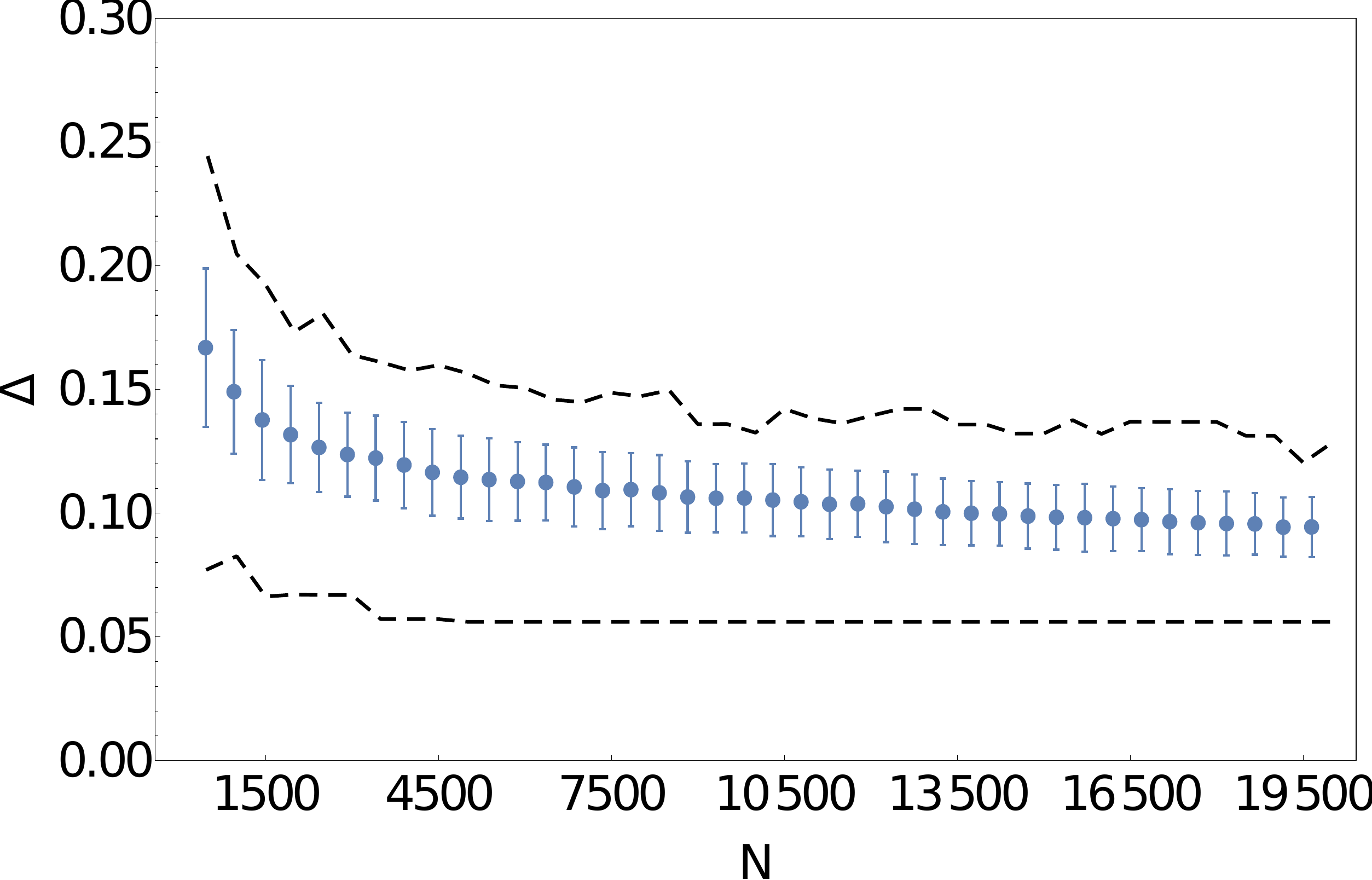}}
\caption{The change of the energy gap $\Delta$ near $E=0$ with the increasing size of the system $N$ for the minimal RFLs. The values were averaged over 200 realizations. The error bar indicates one standard deviation, and the black dashed lines represent the minimal and the maximal value obtained in 200 realizations. The numerical values were obtained via the direct diagonalization.}
\label{energy_gap_fig}
\end{center}
\end{figure}

\begin{figure}[tb]
\begin{center}
\resizebox{0.8\columnwidth}{!}{\includegraphics{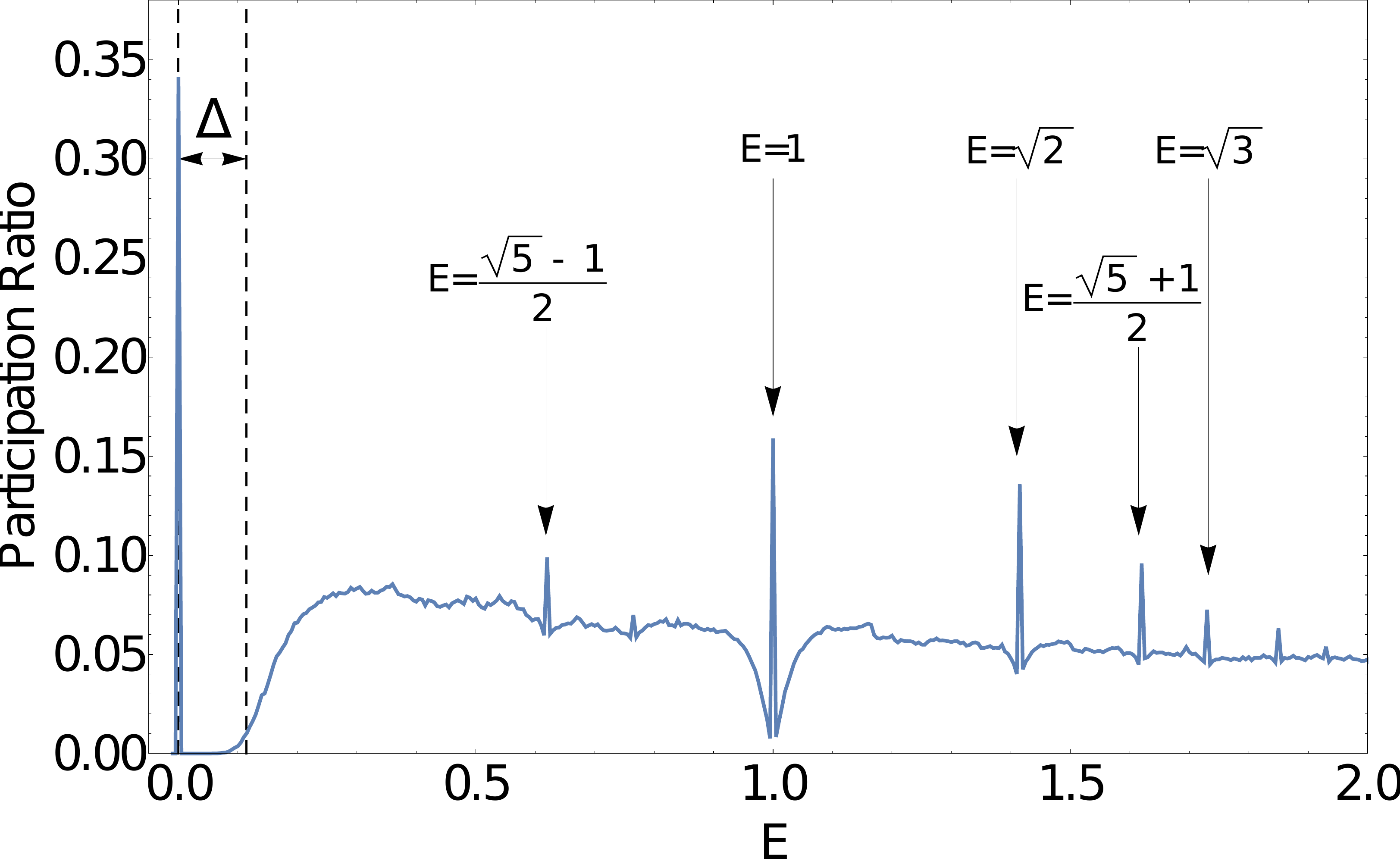}}
\caption{A plot of the participation ratio $PR(E)$ of the minimal RFLs (averaged over 2000 realizations). The participation ratio $PR$ confirms the existence of the energy gap $\Delta$. Also, an additional structure is revealed - we observe some very narrow superlocalization resonances for a discrete set of degenerate energies, where $E_r=0,\pm\frac{\sqrt{5}\pm1}{2}, \pm1, \pm\sqrt{2}$ are the most dominant. For the resonant eigenenergies $E_r$ the system reduces to a few very small clusters, see Fig.~\ref{superlocalization_fig}. Note that the energy gap $\Delta$ is averaged over many realizations and a single realization value might differ, see Fig.~\ref{energy_gap_fig}. Due to this fact, the averaged $PR(E)$ is smeared around $E=\Delta$.} 
\label{pr}
\end{center}
\end{figure}

In Fig.~\ref{phase_diag} we can observe a  peculiar narrow energy gap $\Delta$ around $E=0$ for $d_s \lesssim 1.4$, which eventually closes up when the spectral dimension is being increased. We find that the value of energy gap $\Delta$ for the minimal RFLs (i.e. $d_s=1.33\pm0.03$) of length $N=5000$ is 
\be\label{energy_gap_value}
\Delta = 0.114\pm 0.017. 
\ee
The energy gap \eqref{energy_gap_value} decreases slightly in larger systems, however, it seems that $\Delta$ survives in the thermodynamical limit, see Fig.~\ref{energy_gap_fig}. In order to illustrate the energy gap more clearly we show in Fig.~\ref{pr} the participation ratio $PR(E)$
\be
PR(E) = \sum_{(i,j)} \Big|\langle (i,j) | \psi(E)\rangle  \Big|^4,
\ee
where $ |\psi(E)\rangle $ is an eigenstate corresponding to an energy $E$ and  $|(i,j) \rangle$ is a state localized at a lattice site denoted by $(i,j)$  in the 2D Euclidean space. The participation ratio is yet another measure of the localization \cite{Krame1993,VanTiggelen1999}. That is, the inverse of $PR$ estimates a number of fractal points on which an eigenstate is localized on.

\begin{figure}[tb]
\begin{center}
\resizebox{0.8\columnwidth}{!}{\includegraphics{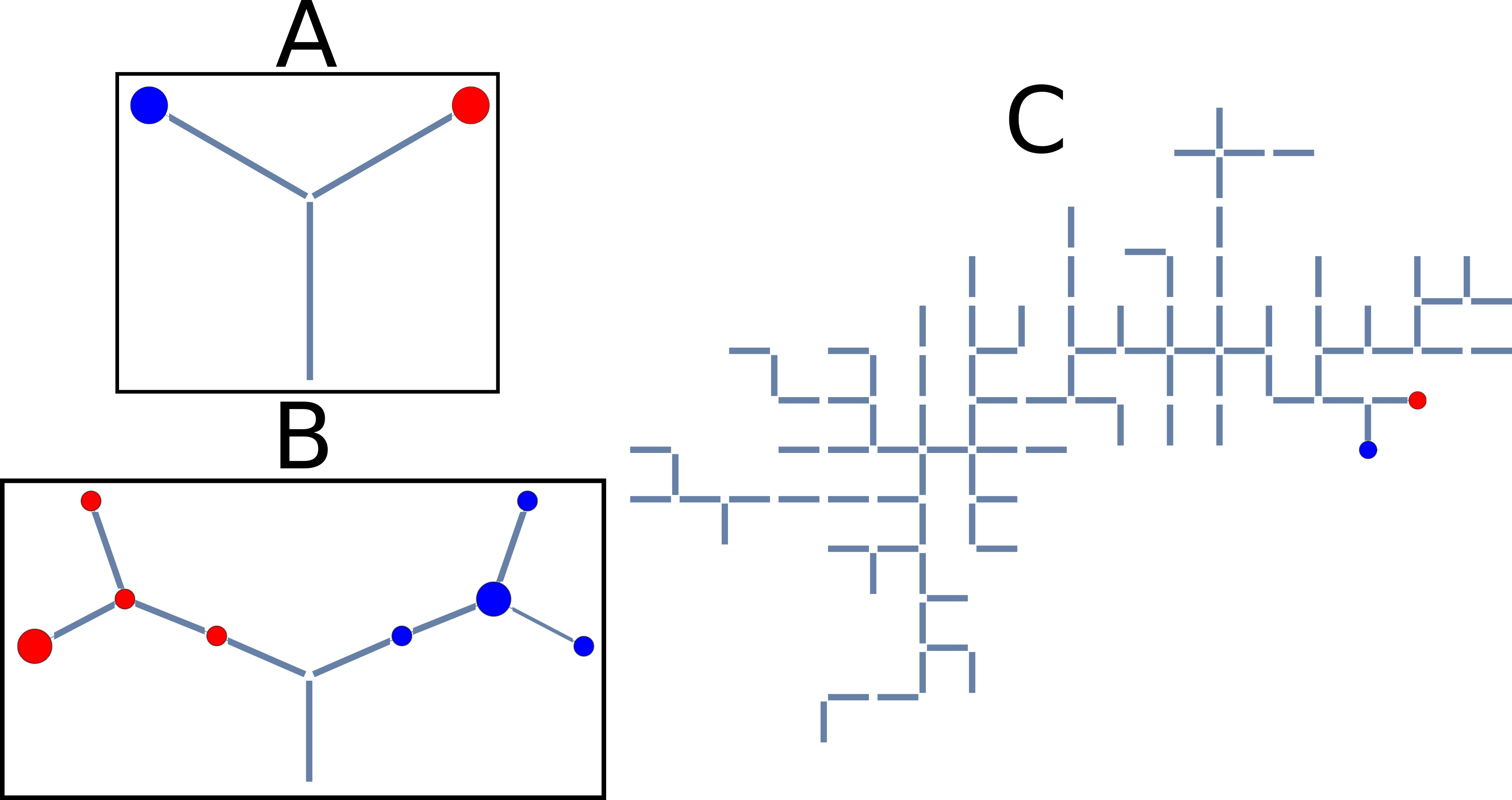}}
\caption{(color online) A sample building blocks responsible for the superlocalization resonances: an eigenstate of a 4-site block  corresponding  to $E=0$ (a) and an eigenstate of a 10-site block related to  $E=\sqrt{3}$ (b). A dot represents a non-zero value of an eigenstate on a given sites $\psi_i$: a dot's size scales with $|\psi_i|$ and red/blue color represents plus/minus sign of $\psi_i$. Notice that a state of block  (a) has the same energy as an eigenstate of a lattice in panel  (c) (connecting empty sites to an empty leg of block (a) does not change its energy). }
\label{superlocalization_fig}
\end{center}
\end{figure}

What is the most striking in Fig.~\ref{pr} is the emergence of peaks that are related to superlocalization resonances also observed in the QP model \cite{Schubert2005}. The resonances appear for a discrete set of energies $E_r$ and they are dominant for $E_r=0,\pm\frac{\sqrt{5}\pm1}{2}, \pm1, \pm\sqrt{2}$. The presence of the resonances is not visible in Fig.~\ref{phase_diag} because in Eq.~\eqref{ri_eq} the degenerate levels are discarded to avoid divergence. The name \emph{superlocalization} stems from the fact that the eigenstates localize on  very small (a few lattice site) disjoint clusters. For example, a zero energy state can be localized on two sites only, as long as a certain building block appears on the lattice boundary. In Fig.~\ref{superlocalization_fig} (a) we see a block with one vertex and four lattice sites. Note, that its zero energy eigenstate has non-zero values on two sites only and the other two sites form an \emph{,,empty leg''}. Now notice, that a structure in Fig.~\ref{superlocalization_fig} (c) must have similar zero energy eigenstates  because connecting empty lattice to an empty leg  of a block like in Fig.~\ref{superlocalization_fig}(a) does not change  a zero energy eigenstate localized on two sites. Therefore, if a lattice geometry allows  for small blocks (like e.g. in panel (a) or (b) of Fig.\ref{superlocalization_fig}), then  some eigenstates of the lattice   coincide with those of small blocks and superlocalization resonances emerge. 

If some blocks are frequently occurring in the lattice, the  corresponding superlocalization resonances   can be extremely degenerate. For example, for the minimal RFLs with $d_H=1.75\pm0.02$  about 10\% of eigenstates have zero eigenenergy.  The zero energy manifold is thus extended over a substantial number of lattice points that is related to the appearance of the energy gap in the spectrum. That is, the zero energy manifold is large and other eigenstates with non-vanishing overlap on the manifold must necessary possess different energies.  

The four-site structure shown in Fig.~\ref{superlocalization_fig}(a) has also non-zero energy eigensolutions, for instance corresponding to $E=-\sqrt{3}$. However, such eigenstates do not have an ,,empty leg'' and therefore if the block is connected to an empty big lattice, these eigenstates are disturbed. Nevertheless, it is possible to build a 3-vertex block, see Fig.~\ref{superlocalization_fig}(b), where there are eigenstates corresponding to $E=-\sqrt{3}$ which possess ,,empty legs'' and are not disturbed when the block is attached to a big empty lattice.
Such a 3-vertex structure is far less common in  RFLs, which explains low abundance of  the $E=-\sqrt{3}$ superlocalization resonance (less then 1\textperthousand).    

 \section{Transmission and quantum evolution}\label{seciv}

In this section we investigate  transport properties of a quantum particle on RFLs:  transmission probability through the lattice and evolution of a particle initially localized on a single lattice site. Here, we focus on small systems with 500 lattice sites only because smaller systems are closer to the experimental reality in ultra-cold atomic gases.

\begin{figure}[tb]
\begin{center}
\resizebox{0.8\columnwidth}{!}{\includegraphics{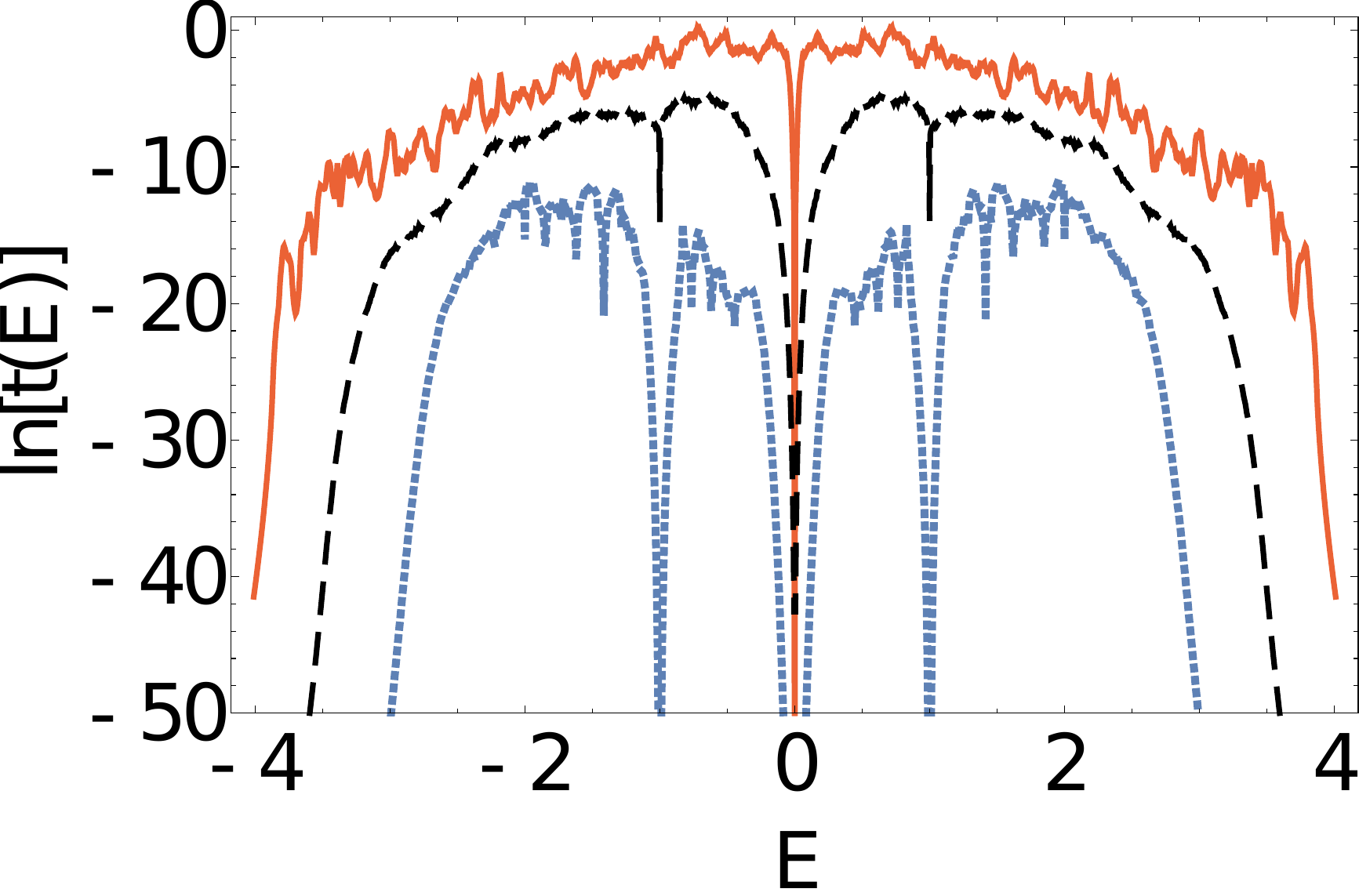}}
\resizebox{0.8\columnwidth}{!}{\includegraphics{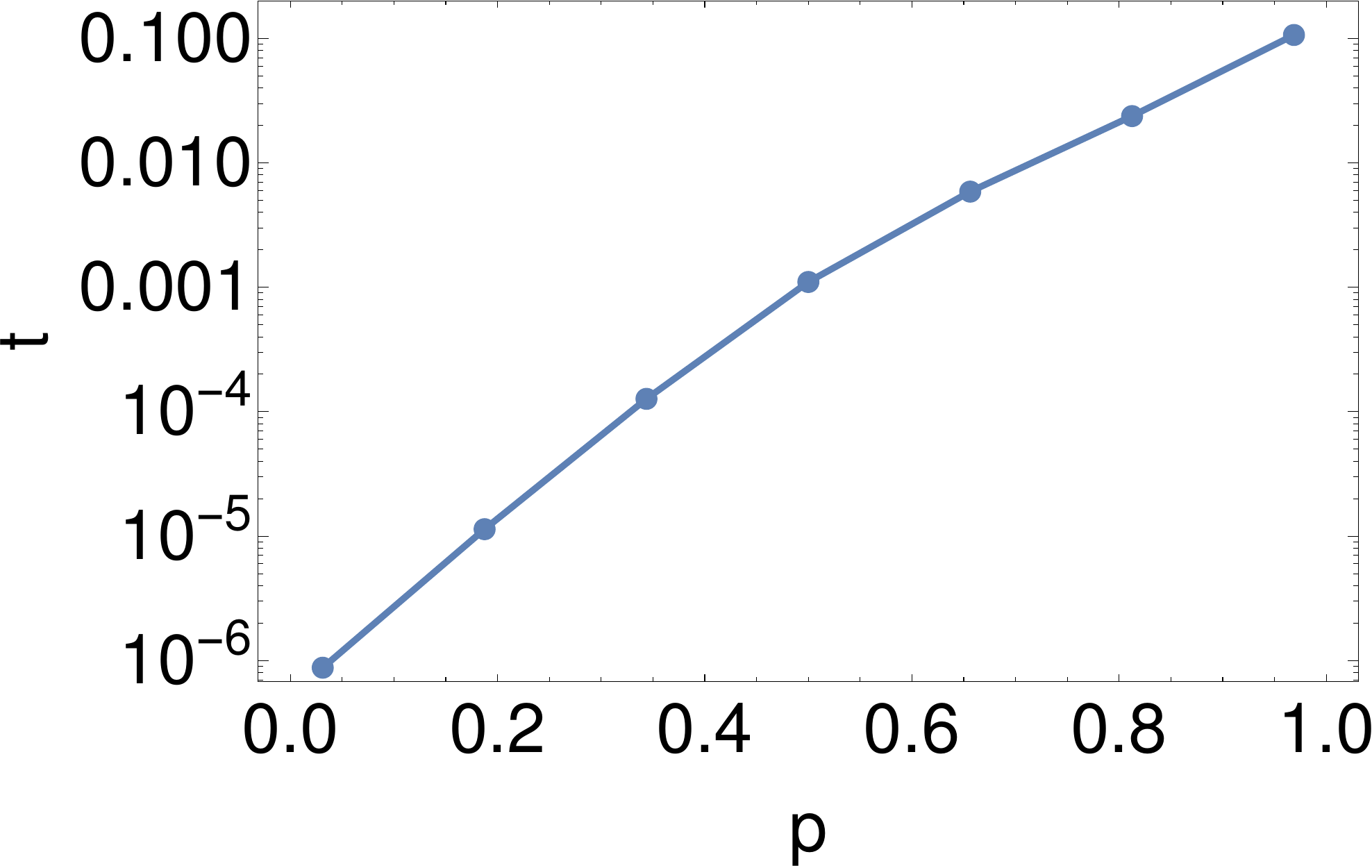}}
\caption{
(color online) The energy dependent transmission probability of a quantum particle between the most distant sites of a lattice of length 500 for $\eta=1$ (top panel) and the same transmission probability averaged    over the corresponding energy spectra (bottom panel). The curves on the top panel were plotted for $p=0.95$ (solid red) $p=0.55$ (black dashed) and $p=0.05$ (blue dotted). We observe that the transition probability strongly increases while we add links to the lattice, which is in the agreement with Fig.~\ref{phase_diag}. Also, one can  distinguish the termination of transport for $p=0.05$ around $E=0$ which corresponds to the energy gap in the system and for the resonant energies $E_r$ (most critically for $E=\pm1$).   The data were averaged over 2000 realizations.}
\label{transmisja_fig}
\end{center}
\end{figure}

\begin{figure}[bt]
\begin{center}
\resizebox{0.9\columnwidth}{!}{\includegraphics{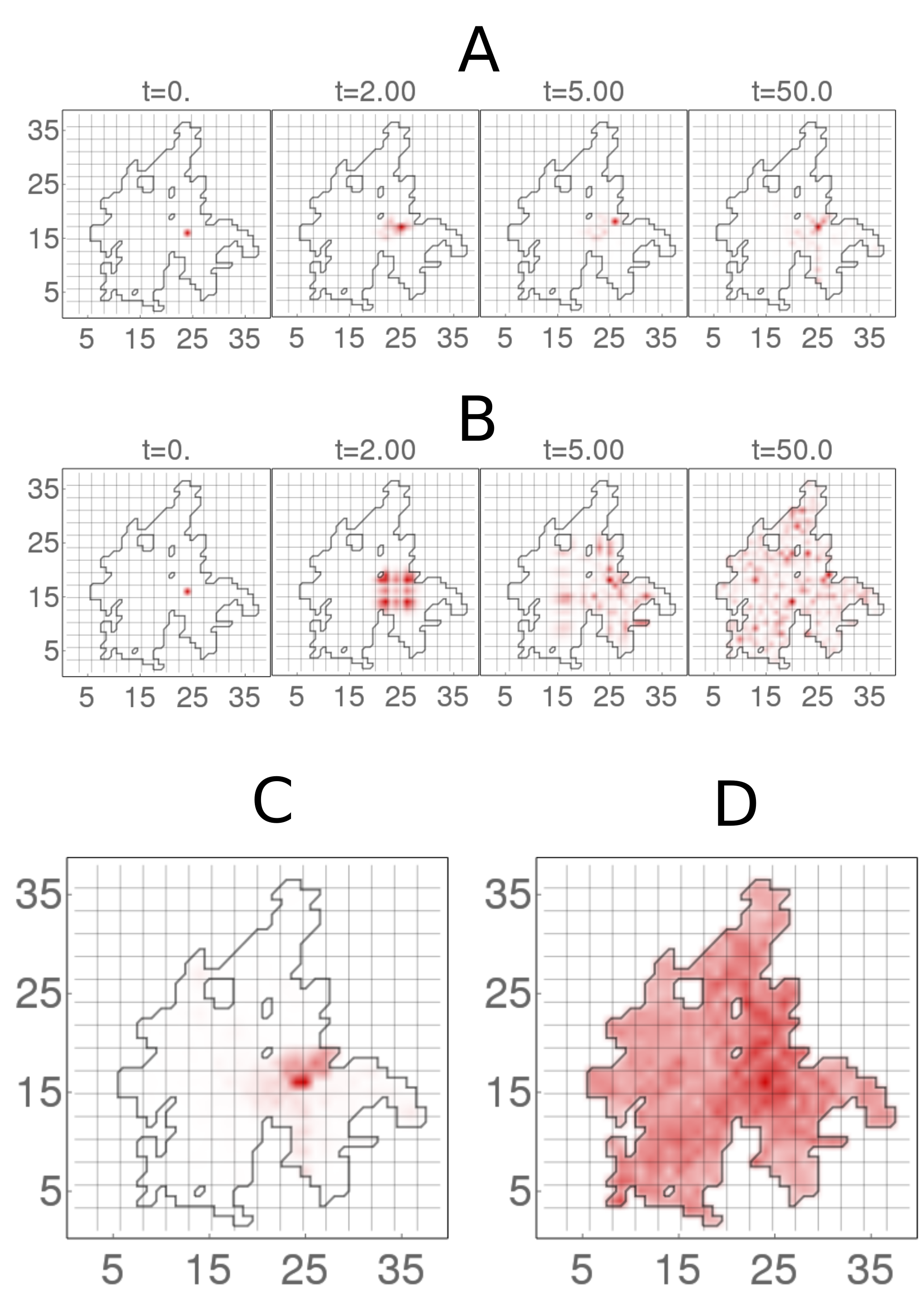}}
\caption{(color online) The evolution of a quantum particle in a sample RFL lattice for the two extreme cases: the minimal number of links (a) and the maximal number of links (b) in a given fractal geometry. We chose a system of 500 lattice sites generated for $\eta=1$, Eq.~\eqref{probability}. The initial state was localized on a single lattice site. Panels (a) and (b) present the probability densities of finding a particle for different evolution times.  Panels (c) and (d) present the time averaged densities for $50<t<150$.} 
\label{ewolucja_fig}
\end{center}
\end{figure}

The transmission probability from site $r$ to site $r'$ of a~quantum particle with the energy $E$ is defined \cite{Krame1993} as:
\be
t(r,r',E)= \left\langle \left| \langle r |G^+(E)| r'\rangle \right|^2 \right\rangle,
\ee
where $\langle ..\rangle$ denotes average over different realizations of RFL,  $G^+(E)=\lim_{\eta\rightarrow 0_+}\left(E+i\eta\,-H\right)^{-1}$ 
is the retarded one-particle Green's function \cite{Economou2006}. We plot the transmission probability throughout considered lattices, i.e. between the most distant sites, in Fig.~\ref{transmisja_fig}. The top panel presents the dependence of the transmission probability on energy for different spectral dimension of RFLs whereas  in  the bottom panel there are  transmission probabilities averaged over entire energy spectra for different number of links in the system. In an  agreement with the results presented in Fig.~\ref{phase_diag}, we observe  a drastic reduction (5 orders of magnitude) of  the transport while  decreasing the number of links in the system (bottom panel). Furthermore, we can see a number of strong dips in the plot of $t(E)$ (top panel), especially for $p=0.05$. These dips correspond  to the energy gap around $E=0$ and the superlocalization resonances for discrete degenerate energies, see Fig.~\ref{pr}. The  most pronounced dips are related to:  $E=0$ (about 10\% of all energy levels correspond to $E=0$), $E=\pm1$  (4\%) and $E=\pm \sqrt{2}$ (1\%).  Note, that  an increase of $p$ (black dashed and red solid curves in the top panel) narrows the dip around $E=0$ significantly down because the gap is disappearing.

Furthermore, the transport properties of a quantum particle can be investigated more directly by solving the time dependent Schr\"odinger equation
\be
i\partial_t \psi_{(i,j)} = -\sum_{i',j'} \psi_{(i',j')},
\ee
cf. Eq.~(\ref{schrod}).

 In panels (a) and (b) of  Fig.~\ref{ewolucja_fig} we present the snapshots of the evolution of a quantum particle  in lattices for the two extreme cases: the minimal and the maximal number of links for a given geometry.  In panels (c) and (d) the time averaged results are shown. Starting from the same initial, fully localized (on a single lattice site) state, we obtain the opposing results: the probability density of finding a particles is either localized around the initial state or explores the whole lattice.

 \section{Conclusions}\label{secv}

We have investigated localization and transport of a quantum particle in lattices with a fractal structure. The lattices consist of points that form a connected cluster. Sites of the lattices are generated so that their fractal (Hausdorff) dimension $d_H$ is controlled. Independently one can control the spectral dimension $d_s$ of the systems by choosing how many nearest neighbor sites are linked to a given lattice point. It allows us to analyze how the localization properties vary with independent changes of the Hausdorff and spectral dimensions.

Analysis of energy level statistics and participation ratio of eigenstates shows that while the localization properties depend very weakly on $d_H$, they change strongly with $d_s$. For the smallest spectral dimension of the systems we observe strong localization of eigenstates. With an increase of $d_s$, eigenstates loose their localization properties and become extended over the entire finite lattices that we consider. Disorder in our systems stems from a non-uniform distribution of lattice points and from their random connections. When $d_s$ approaches $d_H$ all nearest neighbor sites become connected and the randomness is related to the non-uniform distribution of lattice points only. The latter introduces too weak dephasing and eigenstates do not localize.

We observe also eigenstates that are strongly localized on small parts of the random fractal lattices. The smaller part of the fractal, the higher chance for such eigenstates to occur. The zero energy eigenstates can occupy two sites only and consequently they form the largest degenerate manifold. At low spectral dimension they are so many that an energy gap around $E=0$ is created. The presence of strongly localized eigenstates is imprinted in the transport properties of the systems, i.e. the particle transmission probability drops at the corresponding energies. 

\section*{Acknowledgments}
We are grateful to Marcin P\l{}odzie\'{n} for encouraging discussions.

AK acknowledges a support of the National Science Centre, Poland via project DEC-2015/17/N/ST2/04006. KS acknowledges a support of the National Science Centre, Poland via project No.2015/19/B/ST2/01028.



\end{document}